%% file: main.arxiv.tex
\documentclass{preprint}

\usepackage{tabularx}
\usepackage{courier}

\usepackage{xcolor}
\usepackage{booktabs}
\usepackage{graphicx} % Required for inserting images
\usepackage{pifont}
\usepackage{float}
\usepackage{pdfpages}

\usepackage{amsmath,amssymb,amsfonts}
\usepackage{makecell}
\usepackage{multirow}
\usepackage{longtable}
\usepackage{colortbl}
\usepackage{tabularray}
\usepackage{xltabular}
\usepackage{threeparttable}
\usepackage{paralist}
\usepackage{pdflscape}
\usepackage[title]{appendix}

\usepackage[numbers,comma,sort&compress]{natbib}
\usepackage[colorlinks=true]{hyperref}

\title{Deciphering genomic codes using advanced NLP techniques: a scoping review}
\author[1]{Shuyan Cheng}
\author[1]{Yishu Wei}
\author[1]{Yiliang Zhou}
\author[1]{Zihan Xu}
\author[2]{Drew N Wright}
\author[3]{Jinze Liu}
\author[1,*]{Yifan Peng}
\affil[1]{Department of Population Health Sciences, Weill Cornell Medicine, New York, NY 10065}
\affil[2]{Samuel J. Wood Library \& C.V. Starr Biomedical Information Center, Weill Cornell Medicine, New York, NY 10065}
\affil[3]{School of Public Health, Virginia Commonwealth University, Richmond, VA 23219}
\affil[*]{Corresponding:  \url{yip4002@med.cornell.edu}}

\begin{document}

\maketitle

\begin{abstract}

\textbf{Objectives}:
The vast and complex nature of human genomic sequencing data presents challenges for effective analysis. This review aims to investigate the application of Natural Language Processing (NLP) techniques, particularly Large Language Models (LLMs) and transformer architectures, in deciphering genomic codes, focusing on tokenization, transformer models, and regulatory annotation prediction. This review aims to assess data and model accessibility in the most recent literature, gaining a better understanding of the existing capabilities and constraints of these tools in processing genomic sequencing data.

\textbf{Methods}: 
Following Preferred Reporting Items for Systematic Reviews and Meta-Analyses (PRISMA) guidelines, our scoping review was conducted across PubMed, Medline, Scopus, Web of Science, Embase, and ACM Digital Library. Studies were included if they focused on NLP methodologies applied to genomic sequencing data analysis, without restrictions on publication date or article type. 

\textbf{Results}: 
A total of 26 studies published between 2021 and April 2024 were selected for review. The review highlights that tokenization and transformer models enhance the processing and understanding of genomic data, with applications in predicting regulatory annotations like transcription-factor binding sites and chromatin accessibility. 

\textbf{Discussion}: 
The application of NLP and LLMs to genomic sequencing data interpretation is a promising field that can help streamline the processing of large-scale genomic data while providing a better understanding of its complex structures. It can potentially drive advancements in personalized medicine by offering more efficient and scalable solutions for genomic analysis. Further research is needed to discuss and overcome limitations, enhancing model transparency and applicability.

\end{abstract}


\section{Introduction}\label{introduction}
The vast and complex nature of human genomic sequencing data necessitates advanced computational methods for effective analysis and interpretation. In recent years, the intersection of Natural Language Processing (NLP) and data interpretation has garnered significant interest. Large Language Models (LLMs) and transformer architectures, initially designed for natural language understanding, have shown promise in deciphering the genomic code \cite{intro1}. By converting genetic sequences into computationally interpretable formats and leveraging the sophisticated attention mechanisms of transformers, researchers aim to enhance the accuracy and depth of genomic sequencing analysis \cite{intro2}.

The human genome, composed of over 3 billion base pairs, contains information critical for understanding biological processes and disease mechanisms \cite{human_genome}. Traditional methods like Sanger sequencing, next-generation sequencing (NGS), and alignment-based approaches focus on generating and aligning sequence data but often fall short in interpreting large, complex genomic datasets, particularly for identifying regulatory regions and intricate patterns \cite{Goodwin2016}. NLP and LLMs provide a scalable approach beyond raw sequencing, enabling efficient analysis, the discovery of regulatory regions, and deeper insights into genetic variation~\cite{intro4}.

This literature review explores the application of NLP and LLMs in genomic data processing, focusing on three key areas: tokenization of genomic sequences, utilization of transformer models, and prediction of regulatory annotations. Tokenization involves converting raw genomic sequences into a format suitable for analysis, making the data more accessible for computational models \cite{intro4}. With their advanced attention mechanisms, transformer architectures capture complex contextual relationships within the data, providing deeper insights into genomic structures \cite{intro7}. Finally, predictive modeling uses the preprocessed data to identify critical regulatory elements such as transcription-factor binding sites, enhancer-promoter interactions, chromatin accessibility, and gene expression patterns \cite{intro5,intro6}.

By examining these areas, this review aims to highlight the transformative potential of integrating NLP into genomic 
sequencing research. This integration not only enables scientists to leverage the power of LLM to gain a more convenient and deeper understanding of genomic data but also paves the way for advancements in personalized medicine, where treatments can be tailored based on an individual's genetic makeup. Despite the challenges, including data complexity, model interpretability, and validation, the progress in this field holds significant promise for future breakthroughs in genomics and beyond.

\section{Methods}\label{review-process}

\begin{figure}[htbp]
    \centering
    \includegraphics[width=\textwidth]{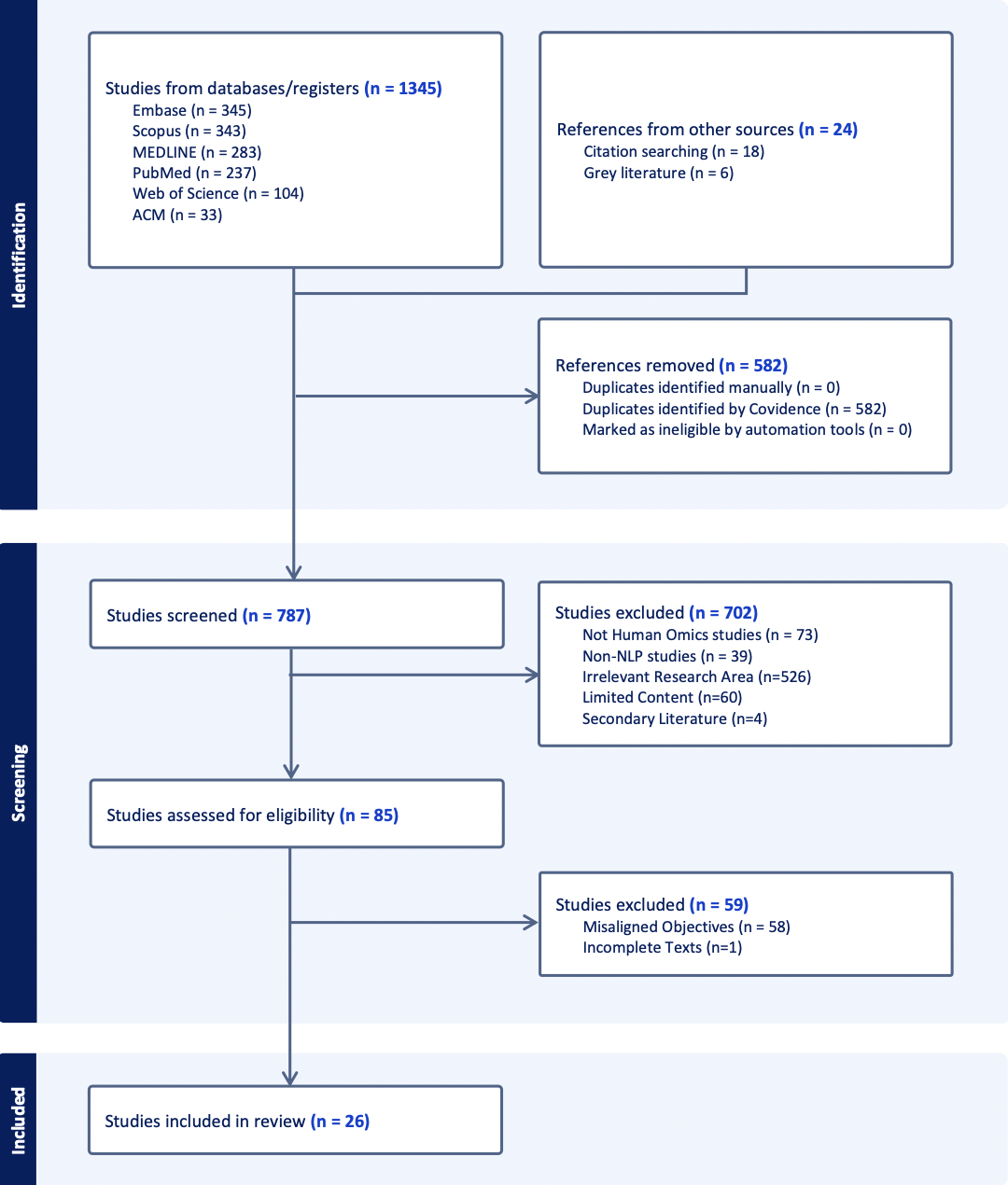}
    \caption{Flowchart of the literature review process according to PRISMA guidelines.} 
    \label{fig:example}
\end{figure}

\subsection{Eligibility criteria}

Our review follows the Preferred Reporting Items for Systematic Reviews and Meta-Analyses (PRISMA) guidelines (\url{https://www.prisma-statement.org/}). The eligibility criteria for the included studies focused on two main areas: NLP and genetic association studies. Studies were included if they specifically addressed applications or methodologies related to NLP in the context of genomic data analysis. No restrictions were placed on publication date or article type, and the primary focus was on studies published in English. A summary of the PRISMA checklist is provided in the Supplementary Materials \ref{sup:prisma}.

\subsection{Information sources}

A comprehensive search was conducted across multiple databases to identify relevant studies. The systematic searches were conducted on Ovid MEDLINE (In‐Process and Other Non‐Indexed Citations and Ovid MEDLINE 1946 to Present), Ovid EMBASE (1974 to present), Scopus, Web of Science, and the ACM Digital Library. The databases searched included PubMed (\url{https://pubmed.ncbi.nlm.nih.gov}), the Institute of Electrical and Electronics Engineers (IEEE) Xplore Digital Library (\url{https://ieeexplore.ieee.org}), Google Scholar (\url{https://scholar.google.com}), and Semantic Scholar (\url{https://www.semanticscholar.org}). The searches were executed in April 2024, with the most recent search conducted in June 2024 to ensure the inclusion of the latest available studies. Important references identified during the searches were also tracked for further examination.

\subsection{Search strategy}

Our search strategy was designed to maximize coverage using broad combinations of keywords and related terms, ensuring a comprehensive capture of relevant literature. It included both controlled vocabulary and free-text terms related to ``natural language processing" and ``genetic association studies." The strategy incorporated keywords and phrases such as ``natural language processing," ``large language model," ``NLP," ``LLM," ``data mining," ``genomic association studies," ``polymorphism," ``SNP," ``token," ``transformer," ``BERT," and ``regulatory annotations." Full details of the search strategies for each database are provided in the Supplementary Materials \ref{sup:search strategies}. 

\subsection{Study selection}

The study selection process involved two stages of screening. Initially, two independent researchers screened the abstracts of all identified articles for relevance to the topics of natural language processing and genetic association studies. Abstracts deemed appropriate were then subjected to a full-text review by the same two researchers to confirm their eligibility for inclusion. The screening was facilitated using Covidence, a web-based tool designed to streamline systematic reviews \cite{covidence}. 

The initial phase led to the exclusion of 702 studies based on the following key exclusion criteria: 1) 73 studies were excluded for not using human omics data 2) 39 studies were excluded because they did not employ natural language processing methods 3) 526 studies were excluded for their irrelevance, particularly those not involving specific NLP downstream tasks or human-omics data 4) 60 studies were excluded for providing insufficient content such as conference proceedings or any submissions that did not provide full-text articles 5) 4 studies consisted of secondary literature, such as survey and review papers. In the full-text screening phase, 59 additional studies were excluded due to misaligned objectives or incomplete texts, where an abstract suggested relevance. Still, the full text lacked sufficient content for relevant analysis.

\section{Results}\label{results}

A total of 26 studies published between 2021 and April 2024 met the inclusion criteria and were selected for final discussion. Table \ref{tab:nlp_genomic} summarizes the main findings.

\begin{landscape}
{%
\tiny\tabcolsep=2pt
\input{table/table1.r0}}
\end{landscape}

\subsection {Preprocessing and Modeling}

Preprocessing genomic sequencing data is crucial before predictive modeling can be applied. This involves converting the raw genomic sequences into a format that computational models can understand, making the complex genetic data more accessible. The preprocessing steps include tokenization, which breaks text into manageable sub-word units. Subsequently, advanced architectures like transformers are utilized during the modeling phase to capture intricate dependencies and patterns in the data.

\subsubsection{Tokenization of Genomic Data for LLMs}

Tokenization is the first step in preprocessing genomic sequencing data for LLMs, which can capture biologically significant patterns such as promoter elements like TATA or CAAT boxes. It involves breaking down the sequences into smaller, manageable, and interpretable units that can be fed into computational models. K-mers is the most widely used tokenization method among the studies reviewed, and it is consistent with common practice in genomic research. Biological functions are often determined by short patterns in DNA sequences, such as motifs and binding sites, making k-mers suited to capture these. Furthermore, many studies are built on top of existing pre-trained models and follow the tokenizer used in the original model. For example, several studies build on DNABERT \cite{1027,174} and employ k-mers accordingly. Beyond k-mers, other tokenization methods are also applied. For instance, Hossain et al. approach the problem of CpG island detection as named entity recognition (NER) and use a BPE tokenizer \cite{1045,946}.

\paragraph*{K-merization:} K-merization is a method in bioinformatics that breaks down DNA sequences into smaller, overlapping segments of fixed length, known as k-mers, where `k' represents the number of nucleotides in each segment \cite{Huang2019}. For instance, in DNABERT, applying k-mers with a value of 3 on a sequence like "ACTGACTGAC" results in tokens such as ["ACT", "CTG", "TGA", "GAC"]. Several studies have effectively utilized k-mer tokenization for sequencing data processing. Wang et al. applied $k=1, 3, 5$ to split DNA sequences into k-mers, treating them as words in a natural language \cite{714}. Additionally, Ji et al. employed various k-mer lengths (3, 4, 5, 6) in training and fine-tuning DNABERT models to understand the DNA sequences and predict regulatory elements \cite{1101}. DNABERT was trained using the Bidirectional Encoder Representations from Transformers (BERT) model \cite{devlin2019bert-t}, which is developed to create deep bidirectional representations by preprocessing text without labels, considering the context from both the left and the right at every layer. Similarly, Zeng et al. introduced a custom corpus and tokenizer using 6-mers with taxonomic lineage descriptions and aimed to predict the DNA methylation sites ($F1=0.95$) \cite{656}. Moreover, An et al. used 6-mers to pre-train human genome data and fine-tune downstream data, getting the moDNA model, which is designed for promoter prediction and transcription start site (TSS) detection ($F1 = 0.862$) \cite{1000}.

\paragraph*{Byte-Pair Encoding (BPE):}

BPE is a tokenization method that iteratively merges the most frequent pairs of characters in a text to create subword tokens, allowing for efficient handling of rare and unseen words \cite{bpe_huggingface}. When BPE is applied to the sequence ``ACTGACTGAC," it may yield tokens like \texttt{["ACTG", "ACTG", "AC"]}. Hossain et al. utilized BPE \cite{bpe_huggingface} to tokenize the DNA sequences \cite{1045}. The tokenized data were then applied to three models, each having a parameter size of 66 million. DistilBERT was set as the benchmark, and Conditional Random Fields (CRF) \cite{lafferty2001conditional-r}, and Attention Mask was added to each layer to detect CpG islands in DNA sequences. This methodology aims to predict the promoter regions and identify epigenetic causes of diseases. The F1 scores for these three models are 0.718, 0.726, and 0.735, respectively.

In 2023, Hossain et al. refined this approach by pre-training the BERT model and CRF layer (parameter size = 110M) on a large sequencing dataset with 142,325 CpG islands and fine-tuning on a smaller dataset \cite{946}. Eventually, they achieved an F-1 score of 0.834.

\paragraph*{Fixed Nucleotide Tokenization:}

Fixed Nucleotide Tokenization is a method that segments DNA sequences into fixed-length nucleotide fragments. This technique differs from K-merization, primarily because it does not allow overlapping segments between the fragments, often treating these fragments as distinct ``tokens" or ``words" in the context of deep learning and NLP models. Using FNT with 3-nucleotide groupings on ``ACTGACTGAC," for example, produces tokens like \texttt{["ACT", "GAC", "TGA", "C"]}. For example, Le et al. divided each DNA sequence into 81-base pair fragments and treated each fragment as a token, ensuring each fragment was treated as an independent token without any overlap \cite{190}. This fragment-based tokenization helps maintain sequence context over a fixed window size. They aimed to identify both promoters and non-promoters and their activities. In this study, researchers proposed the latest pre-trained BERT model, which eventually got promoter identification and strength identification, achieving an accuracy of over 0.8. Another study presented a novel method that combines BERT and 2D convolutional neural networks (CNNs) to predict DNA enhancers \cite{232}. During the training process, each nucleotide was transformed into a contextualized word embedding vector of size 768. These vectors, which represent fixed-length sequences, were then passed to a CNN for further analysis. The model demonstrated superior performance by training on the iEnhancer-2L dataset, achieving an accuracy of 0.756 and a sensitivity of 0.8. In addition, Rajkumar et al. presented a transformer-based model named DeepViFi that tokenized each nucleotide base (A, C, G, T, N) individually, rather than using k-mers or sub-sequences \cite{1064}. This approach leveraged a random forest classifier to identify viral reads in cancer genomes, particularly focusing on the Human Papillomavirus (HPV), and a LightGBM model to classify these viral reads into specific subfamilies. The results showed that DeepViFi achieved a high precision-recall AUC of 0.94 in detecting and classifying HPV reads, demonstrating its effectiveness in this domain.

%\paragraph*{Gene Names and Binned Values:} 

%Cui et al. \cite{308} applied gene names and binned values for tokenizing expression levels in single-cell RNA sequencing (scRNA-seq) data. This method enabled the handling of gene expression levels directly and transformed complex expression patterns into a manageable format, which is particularly useful in single-cell studies where the expression of individual genes needs to be analyzed across thousands of cells.

\subsubsection{Transformer Architecture for Genomic Sequencing Data}

Once the data is tokenized, the next step involves using transformer architectures to capture the complex, contextual relationships within them. Transformers are highly effective due to their attention mechanisms, which allow models to focus on different parts of the sequence simultaneously.

Since most studies are phrased as prediction or classification problems, BERT and its variants are often used as feature extractors, with an additional classifier added on top of that. In this section, we will mostly focus on the BERT and transformer components. For sequencing labeling tasks, the transformer is directly used \cite{212,308}. Additionally, some models incorporate CNNs within transformer-based architectures to enhance local feature extraction. CNNs are particularly effective for capturing motifs and short sequence patterns, which are essential for refining sequence-level predictions within broader transformer architectures.

\paragraph{BERT and Variants:}

After tokenizing the DNA sequences, Wang et al. pre-trained and fine-tuned a BERT model to predict 5-methylcytosine (5mC) sites and identify DNA enhancers \cite{714}. The BERT-5mc model derived an accuracy of 0.933. The DNABERT models by Ji et al. on different k-mers have 110M parameters and indicate the F1 values over 0.9 for all `k' \cite{1101}. Luo et al. introduced TFBert, a model based on the BERT architecture specifically designed to predict DNA-protein binding sites \cite{174}. The model was derived by initializing with the DNABERT pretraining model and then performing task-specific pretraining on a large dataset of 690 ChIP-seq datasets consisting of various DNA-protein binding data. The model tokenized DNA sequences into k-mers, treating them as words in the context of a language model, allowing it to capture the context of DNA sequences effectively. The primary goal of TFBert is to improve the accuracy and robustness of DNA-protein binding predictions, especially in cases where the datasets are small or medium-sized. The results demonstrated that TFBert \cite{174} achieved state-of-the-art performance, outperforming other existing models, with an average AUC of 0.947, making it a valuable tool for various biological sequence prediction tasks.

\paragraph{Transformer Encoder Blocks:}

Besides BERT, several studies only utilized vanilla transformer encoder blocks or modified versions, which refer to the original transformer architecture with basic attention and feed-forward layers, without additional layers or task-specific pretraining. Unlike models like BERT, which are optimized with masked language modeling and specialized layers for specific downstream tasks, vanilla transformers typically lack these enhancements and may require additional tuning to process complex genomic data effectively. Pipoli et al. proposed a transformer-based model called Transformer DeepLncLoc to process the DNA sequences into a more compact and informative representation using a k-mer approach combined with word2vec embedding \cite{507}. It was specifically designed to process gene promoter sequences and predict the abundance of mRNA, managing the task as a regression problem. The transformer model was then used to analyze these embedded sequences, and its performance was compared against other models like LSTM DeepLncLoc and a convolutional model called DivideEtImpera, which utilized CNN layers to capture local sequence features. Including CNNs in this context helped capture motifs and short patterns within the sequence, providing an advantage in prediction accuracy. Roy et al. introduced a novel approach for masked language modeling (MLM) in gene sequences, specifically focusing on improving the efficiency and performance of transformer-based models like DNABERT and LOGO \cite{902}. The paper presented the GENEMASK model (parameter size = 110M), derived by applying a Pointwise Mutual Information (PMI)-based masking strategy to gene sequences. This strategy identifies and masks the most correlated spans of nucleotides, as opposed to the random masking strategy used in traditional models. The GENEMASK model aims to enhance the learning process by making it more challenging and reducing the pretraining time while improving accuracy, particularly in few-shot settings where training data is limited. The results demonstrated that GENEMASK significantly outperforms the baseline models in several gene sequence classification tasks, showing better accuracy ($0.898 \pm 0.005$) and ROCAUC ($0.962 \pm 0.002$), especially in low-resource scenarios.

\paragraph{Advanced Attention Mechanisms:}

Advanced attention mechanisms are specialized features within transformer architectures that enhance the model's capacity to identify and capture complex relationships within genomic data. Wang et al. presented a novel deep learning model called MTTLm(6)A, designed to predict N6-methyladenosine (m6A) sites on mRNA at base resolution \cite{859}. The model employed a multi-task transfer learning approach, leveraging information from related tasks to improve m6A site prediction. The primary model is an improved transformer architecture, fine-tuned using datasets from various species to enhance its generalization capabilities. The results demonstrated that MTTLm(6)A outperformed other state-of-the-art models in terms of prediction accuracy and efficiency. In another study, Wang et al. proposed MSCAN, a deep learning framework designed for RNA methylation site prediction, mainly focused on identifying various types of RNA modifications \cite{48}. The model incorporated multi-scale self- and cross-attention mechanisms to capture both local and long-range dependencies in RNA sequences. Using different input sequence scales, the model effectively captured the complex contextual relationships crucial for accurate methylation site prediction. MSCAN outperformed existing models in predicting 12 different RNA modifications.

\subsection{Predicting Regulatory Annotations}

After preprocessing and deriving embeddings through the transformer, the tokenized and transformed genomic data can predict various regulatory annotations. These predictions include identifying transcription-factor binding sites, enhancer-promoter interactions, chromatin accessibility, and gene expression patterns. Many studies have demonstrated significant success in these predictive tasks, and the performance of selected models is summarized in Table \ref{tab:nlp_results}. Our selection is based on studies that employed predictive models and reported at least one performance metric, ensuring high standards of empirical validation. 

\paragraph{Methylation and Epigenetic Sites Detection:} 

Recent studies in NLP applied to genomic data have focused on predicting various DNA methylation sites, such as 5-methylcytosine \cite{714}, N6-adenine, N4-cytosine, and 5-hydroxymethylcytosine \cite{656}, which are crucial for understanding epigenetic modifications and their roles in gene regulation. These efforts are integral to advancing our understanding of gene expression regulation, epigenetic modifications, and their broader implications in health and disease. Future research may further explore other underrepresented epigenetic modifications to expand the applications of NLP in this critical area.

\paragraph{Transcription-Related Predictions:} 

Other research efforts are dedicated to genome embedding, promoter prediction, and transcription factor binding site prediction. These goals are essential for mapping the regulatory landscapes of genomes, which is a critical step in understanding how genes are controlled and expressed in different conditions. In parallel, detecting CpG islands in DNA sequences has been a focal point \cite{1045, 946}, with models designed to predict promoter regions and identify epigenetic causes of diseases. This line of research aims to unravel the genetic factors that contribute to various diseases, providing insights that could lead to new therapeutic strategies.\par

\paragraph{Post-Transcriptional Interaction Predictions:} 

Another important application of NLP in genomic research is predicting interactions between various non-coding RNAs and their targets, which play crucial roles in post-transcriptional regulation. Specifically, studies have focused on predicting circRNA-miRNA interactions \cite{12, 69}, and miRNA-mRNA interactions \cite{40}. These interactions are critical for understanding the complex regulatory networks that control gene expression after transcription and influence processes such as mRNA stability, translation, and degradation. By accurately predicting these interactions, NLP models can provide valuable insights into the mechanisms of gene regulation at the post-transcriptional level, which has significant implications for understanding diseases, developing biomarkers, and designing targeted therapies. This section also includes identifying RNA methylation sites, such as m6A \cite{859} and m7G \cite{34}, which are vital for understanding RNA biology and its impact on gene expression.

\paragraph{Cancer Research and Oncology:}

 In the context of cancer research, NLP models have been applied to several specialized tasks aimed at improving cancer diagnosis and treatment. These tasks include predicting optimized potential anti-breast cancer therapeutic target genes \cite{986}, enhancing tumor type classification \cite{1027}, and providing machine learning models capable of handling omics data. Moreover, models have been developed to investigate the reusability and generalizability of cell-type annotation in single-cell RNA sequencing data, which is crucial for understanding tumor heterogeneity, as well as the regulatory mechanisms that control gene activity within cancerous tissues. Additionally, models have been developed to investigate the reusability and generalizability of cell-type annotation in single-cell RNA sequencing data, which is crucial for deepening our knowledge of cellular functions and the complex interactions that drive cancer progression. Another significant goal is the detection of oncoviral infections in cancer genomes using transformers \cite{1064}, a critical step in understanding the role of viral integration in cancer development and progression. Collectively, these applications of NLP in oncology highlight its potential to revolutionize cancer research by providing more precise diagnostic tools, therapeutic strategies, and insights into the multi-layered biological data that underpin cancer biology. 
 
\paragraph{Emerging Directions:}
 
Some goals have been explored less frequently in the literature but hold significant potential for future research. For instance, using NLP models to analyze gene-disease associations across vast biomedical literature can help uncover novel genetic risk factors and pathways associated with complex diseases. Similarly, chromatin accessibility prediction, which involves identifying regions of the genome that are open and accessible to transcription factors, is crucial for understanding gene regulation \cite{979}. These emerging research directions could guide future studies, encouraging researchers to explore these underrepresented yet critical areas, ultimately expanding the applications of NLP in genomics and enhancing our understanding of complex biological processes.

\subsection{Data Diversity and Accessibility}

Upon reviewing recent research on the application of NLP in genomic sequencing data interpretation, a distinct trend in the types of data used becomes evident. DNA sequences dominate the landscape, highlighting their frequent use in NLP-driven genomic analyses. There is also a growing incorporation of RNA sequences and multi-omic data, reflecting a shift toward more diverse and comprehensive datasets.

Additionally, specialized data types, such as structural data from 3D Magnetic Resonance Imaging (MRI) and genomic variations like Single Nucleotide Polymorphisms (SNPs) and Copy Number Variations (CNVs), along with advanced sequencing technologies like Nanopore sequencing, are also being integrated. This broadens the scope of analysis within NLP applications, as these data types provide complementary information beyond traditional sequence analysis. This trend indicates an ongoing expansion in the variety of genomic and multimodal data utilized in NLP for genomics.

Among the studies, most datasets are publicly accessible, with a few studies having limited or request-based access for specific subsets. It fosters inclusivity and sustainable development in integrating genomic data with NLP, enhancing collaboration and progress in this rapidly evolving field.\par

\paragraph{Computational Resources Requirements:}
Advanced NLP techniques are notoriously known for high hardware requirements. However, the actual computational demands for application vary significantly depending on the extent of model training involved. Pretraining models like BERT or large language models (LLMs) from scratch requires significant computational resources. For example, Ji et al. trained DNABERT for 25 days using 8 GPUs \cite{1101}, while Zhang et al. trained on 6,000 GPUs \cite{40}. In contrast, using existing pretrained models as feature extractors and building a classifier such as XGBoost \cite{190,69} or small neural networks \cite{1027,12} on top have minimal requirements. Fine-tuning or continuously pretraining from a publicly available model lies between these extremes \cite{656,174}. In addition, some studies intentionally consider resource constraints in model design and training processes. For example, Roy et al. stopped training at 10,000 steps due to resource limitations and diminishing marginal returns to training \cite{902}. Furthermore, Wang et al. designed a small architecture (a two-layer transformer) to fit into low-resource environments \cite{143}.

%place for insert table 
%\includepdf[pages=-,width=\paperwidth,pagecommand={},fitpaper=true]{table/Figure1.pdf}
% \setlength{\extrarowheight}{3pt}
% \begin{table}[ht]
% \caption{An overview of NLP application using big genomic data}
% \label{tab:nlp_genomic}
% \end{table}
% \includepdf[pages=2-4,width=\paperwidth,pagecommand={},fitpaper=true]{table/NLPforGene_table.pdf}

\begin{table}
\centering
\caption{Detailed Results of Selected NLP Applications in Genomic Sequencing Data Analysis}
\label{tab:nlp_results}
\resizebox{\textwidth}{!}{
\begin{tabular}{lrrrrrrr}
\toprule
\textbf{Model} & \textbf{Accuracy} & \textbf{F1} & \textbf{MCC} & \textbf{ROCAUC} & \textbf{Specificity} & \textbf{Precision} & \textbf{Recall}\\
\midrule
BERT-5mC \cite{714} & 0.933 & - & 0.656 & 0.966 & 0.938 & - & 0.872 \\
DNABERT \cite{1101}\textsuperscript{a} & 0.965 & 0.965 & - & 0.930 & - & - & - \\
SETOMIC \cite{1027} & 0.950 & 0.921 & - & 0.997 & - & 0.945 & - \\
SETQUENCE \cite{1027} & 0.475 & 0.359 & - & 0.910 & - & 0.375 & - \\
BERT-CNN \cite{232} & 0.756 & - & 0.514 & - & 0.712 & - & 0.800 \\
TFBERT \cite{174} & 0.880 & 0.880 & 0.762 & 0.947 & - & 0.882 & 0.880 \\
IGnet \cite{143} & 0.838 & 0.824 & - & 0.924 & - & 0.875 & 0.778 \\
MuLan-Methyl \cite{656} & 0.948 & 0.950 & - & 0.968 & - & - & 0.979 \\
moDNA \cite{1000} & 0.862 & 0.862 & 0.725 & 0.935 & - & 0.863 & 0.862 \\
DistilBERT+CRF+Attention Mask \cite{1045} & 0.965 & 0.735 & - & - & 0.959 & 0.691 & 0.852 \\
BERT+CRF (with/without) \cite{946} & 0.973 & 0.834 & - & - & 0.962 & 0.780  & 0.897 \\
BERT-Promoter \cite{190} & 0.855 & - & - & - & 0.866 & - & 0.843 \\
DeepViFi (pipeline) \cite{1064} & 0.960 & - & - & 0.94  & - & 0.996 & 1.000 \\
GENEMASK-based \cite{902} & 0.898 & - & - & 0.962 & - & - & - \\
MSCAN \cite{48} & 0.957 & 0.713 & 0.710 & 0.937 & 0.994 & 0.905 & - \\
MTTLm6 \cite{859} & 0.699 & 0.713 & 0.399 & 0.771 & 0.649 & 0.681 & -    \\
BioDGW-CMI \cite{12}\textsuperscript{a} & 0.885 & 0.885 & - & 0.948 & - & 0.885 & 0.885 \\
BCMCMI \cite{69} & 0.832 & 0.836 & 0.667 & 0.904 & - & 0.808 & 0.868 \\
MiTDS \cite{40} & 0.770 & 0.810 & - & - & - & - & 0.960 \\

\bottomrule
\end{tabular}
}
\begin{tablenotes}
\footnotesize
\item[a] \textsuperscript{a}The results represent the best model performance under the tested configurations.
\end{tablenotes}

\end{table}

\section{Discussion}\label{discussion}

Applying LLMs within NLP for genomic data interpretation significantly advances the processing and analysis of complex biological data. This review highlights key areas where these technologies have been effectively utilized, including tokenization techniques, transformer architectures, and the prediction of regulatory annotations. While the progress in these areas is promising, several challenges and opportunities for future research remain.

One of the major challenges in applying NLP and LLMs to genomic data is the inherent complexity of the data itself \cite{challenges}. Genomic sequences contain vast information, making it difficult to capture the full context within a model. This complexity also impacts model interpretability, as the black-box nature of LLMs makes it challenging for researchers to understand how the model arrives at its predictions. A `black-box' model refers to a system where the internal workings are not transparent or easily understood, and training data is obscured or undocumented, making it difficult to trace how specific inputs are transformed into outputs \cite{blackbox}. For instance, while models like DNABERT \cite{1101} have successfully predicted regulatory elements and annotated single-cell RNA data, the pathways and features leading to these predictions are often vague, limiting their utility in clinical settings.

To address this issue, future research should focus on developing methods that enhance model interpretability. Techniques such as attention visualization, feature attribution, and post-hoc analysis can provide insights into which parts of the genomic sequence most influence the model's predictions. By making these models more transparent, researchers and clinicians can gain greater confidence in their use for decision-making in personalized medicine.

Another significant challenge genomic researchers face is the absence of well-established pipelines or guidelines for integrating LLMs into genomic data analysis, such as determining which models best suit different data types, such as DNA/RNA sequences, proteomics, or epigenomics data. In addition, k-mers is still the most popular tokenizer, which might be sub-optimal \cite{dotan2024effect}. Selecting the best tokenizer for relevant tasks needs further investigation. Although LLMs have shown great promise in various applications, their use in genomics is still in its early stages, often requiring ad hoc and highly specialized approaches. Developing systematic pipelines that outline best practices for tokenization, model selection, training, and validation in the context of genomic data is crucial. Such guidelines would standardize the use of LLMs across research groups, ensuring reproducibility and reliability of results. Moreover, these frameworks could make LLMs more accessible to researchers with limited computational backgrounds, helping streamline their adoption in genomics.

One limitation of this study is the constrained scope of the literature review sources, which primarily includes human genome data and only limited exploration of bacterial, viral, and other non-human DNA. Moreover, the study predominantly focuses on cancer when it comes to disease analysis, giving relatively less attention to other disease domains that involve complex DNA interactions, such as neurodegenerative diseases, autoimmune diseases, and genetic disorders. These areas also offer rich opportunities for genomic research and could benefit from applying NLP techniques.

Future models should also consider integrating multimodal data more, such as combining genomic sequences with transcriptomic and proteomic data, and clinical data, such as lab values and diagnoses. Integrating multiple types of genomic data can better reveal complex biological interactions, providing insights into how different layers of biological information interact to drive cellular functions \cite{discussionmulti-omic}. Including clinical data can enhance model predictions by grounding them in real-world patient information, thereby improving clinical relevance and enabling personalized insights. This comprehensive approach can provide a more comprehensive understanding of the regulatory mechanisms governing gene expression and the interplay between different molecular layers. It can also enable models to generate and validate more accurate and biologically meaningful predictions, thereby increasing physicians' confidence in NLP-generated results and promoting the widespread application of NLP-based models.

Ultimately, integrating NLP and LLMs into genomics aims to translate these advancements into practical applications. This includes developing models that can predict individual responses to treatments based on genomic data, identify potential therapeutic targets, as well as provide clinicians with actionable and interpretable insights. As the field progresses, collaboration between computational scientists, geneticists, and clinicians will be essential to ensure that these models are both scientifically valid and clinically useful.

\section*{Funding Statement}

This work was supported by the National Library of Medicine under Award No. R01LM014306.

\section*{Competing Interests Statement}

There are no competing interests to declare.

\section*{Contributorship Statement}

Study concepts/study design: Shuyan Cheng, Jinze Liu, Yifan Peng; manuscript drafting or manuscript revision for important intellectual content: all authors; approval of the final version of the submitted manuscript: all authors; agrees to ensure any questions related to the work are appropriately resolved: all authors; literature research: Shuyan Cheng, Drew N Wright; data interpretation: Shuyan Cheng, Yishu Wei, Yiliang Zhou, Zihan Xu; and manuscript editing: all authors.

\section*{Data Availability}

The data underlying this article are available in the article and in its online supplementary material.

\section*{Abbreviation}

\begin{longtable}{ll}
3D MRI &  Three-Dimensional Magnetic Resonance Imaging \\
5mC & 5-methylcytosine\\
AUC & Area Under the Curve\\
AUC-PR & Area Under the Precision-Recall Curve\\
BCMCMI & Fusion Model for Predicting circRNA-miRNA Interactions\\
BERT & Bidirectional Encoder Representations from Transformers\\
BPE & Byte-Pair Encoding\\
ChIP-seq & Chromatin Immunoprecipitation Sequencing\\
CNN & Convolutional Neural Network\\
CNV & Copy Number Variation\\
CpG & Cytosine-phosphate-Guanine\\
CRF & Conditional Random Fields\\
DeepLncLoc & Transformer Model for Predicting mRNA Abundance\\
DeepViFi & Detecting Oncoviral Infections in Cancer Genomes Using Transformers\\
DNABERT & DNA-specific Version of BERT\\
DNA & Deoxyribonucleic Acid\\
FNT & Fixed Nucleotide Tokenization\\
GENEMASK & Gene Masking Model\\
GRM & Genetic Relationship Matrix\\
IGnet & Imaging Genetic Network\\
K-mer & Sequence of K Nucleotides\\
LLM & Large Language Model\\
MCC & Matthews Correlation Coefficient\\
miRNA & MicroRNA\\
MLM & Masked Language Modeling\\
m6A & N6-methyladenosine\\
m7G & N7-methylguanosine\\
MSCAN & Multi-Scale Self- and Cross-Attention Network\\
MTTLm6A & Multi-Task Transfer Learning for m6A Prediction\\
NGS & Next-generation Sequencing\\
NLP & Natural Language Processing\\
NR & Not Reported\\
PRISMA & Preferred Reporting Items for Systematic Reviews and Meta-Analyses\\
ROCAUC & Receiver Operating Characteristic - Area Under Curve\\
SETQUENCE & Deep Set Transformer-Based
Representations for Cancer Multi-Omics\\
SHAP & SHapley Additive exPlanations\\
SNP & Single Nucleotide Polymorphism\\
TF & Transcription Factor\\
TFBERT & Transcription Factor Binding Site BERT\\
TMSC-m7G & Transformer Model for RNA N7-methylguanosine Site Identification\\
miTDS & miRNA-Target Detection System\\
RNA & Ribonucleic Acid\\
iEnhancer-2L & Dataset for DNA Enhancer Prediction\\
\end{longtable}

\bibliographystyle{unsrtnat}
\bibliography{ref.r0}

\newpage
%\appendix

\begin{appendices}

%\section{Supplementary materials}

\section{PRISMA checklist}
\label{sup:prisma}

This review adheres to PRISMA guidelines to ensure methodological rigor and transparency. Key aspects of the PRISMA checklist addressed in this review include defining clear objectives and eligibility criteria, describing information sources and search strategies, and outlining the study selection and data collection processes. Additionally, we summarize study characteristics, assess the risk of bias, and present the synthesis results and findings. Limitations are discussed, along with an assessment of evidence certainty to contextualize the main findings. Funding sources and conflicts of interest are also disclosed to maintain transparency.

For further details on the PRISMA checklist and specific reporting requirements, please refer to the full PRISMA 2020 checklist available at \url{http://www.prisma-statement.org/PRISMAStatement/Checklist}.

\section{Search Strategies}
\label{sup:search strategies}

\subsection*{Medline (04/12/24)}
\begin{compactenum}
    \item exp Natural Language Processing/
    \item exp Data Mining/
    \item ((natural adj2 language adj2 process*) OR (large adj2 language adj2 model*) OR "NLP" OR "LLM" OR ((data OR text) adj2 (mine* OR "mining")) OR "named entity" OR semantic).ti,ab.
    \item 1 OR 2 OR 3
    \item exp Genetic Association Studies/
    \item exp Polymorphism, Genetic/
    \item (genet* OR genom* OR genot* OR sequenc* OR ((single adj2 nucleotide adj2 polymorph*) OR "SNP*")).ti,ab.
    \item 5 OR 6 OR 7
    \item (token* OR "transformer" OR "bert" OR (regulat* adj2 annotat*)).ti,ab.
    \item 4 AND 8 AND 9
\end{compactenum}

\subsection*{Embase (04/12/24)}
\begin{compactenum}
    \item exp natural language processing/
    \item exp data mining/
    \item ((natural adj2 language adj2 process*) OR (large adj2 language adj2 model*) OR "NLP" OR "LLM" OR ((data OR text) adj2 (mine* OR "mining")) OR "named entity" OR semantic).ti,ab.
    \item 1 OR 2 OR 3
    \item exp genetic association study/
    \item exp genetic polymorphism/
    \item (genet* OR genom* OR genot* OR sequenc* OR ((single adj2 nucleotide adj2 polymorph*) OR "SNP*")).ti,ab.
    \item 5 OR 6 OR 7
    \item (token* OR "transformer" OR "bert" OR (regulat* adj2 annotat*)).ti,ab.
    \item 4 AND 8 AND 9
\end{compactenum}

\subsection*{PubMed (04/12/24)}

\begin{quote}
("Natural Language Processing"[Mesh] OR "Data Mining"[Mesh] OR NLP OR LLM OR "Natural Language Processing" OR "data mining") \\
AND ("Genetic Association Studies"[Mesh] OR "Polymorphism, Genetic"[Mesh] OR genet* OR genom* OR genot* OR sequenc* OR "SNP" OR "SNPs") \\
AND (token* OR "transformer" OR "bert" OR "regulatory annotations")
\end{quote}

\subsection*{Scopus (04/15/24)}

\begin{quote}
( TITLE-ABS-KEY ( ( genet* OR genom* OR genot* OR sequenc* OR ( ( single W/2 nucleotide W/2 polymorph* ) OR "SNP*" ) ) ) ) \\
AND ( TITLE-ABS-KEY ( ( ( natural W/2 language W/2 process* ) OR ( large W/2 language W/2 model* ) OR "NLP" OR "LLM" OR ( ( data OR text ) W/2 ( mine* OR "mining" ) ) OR "named entity" OR semantic ) ) ) \\
AND ( TITLE-ABS-KEY ( ( token* OR "transformer" OR "bert" OR ( regulat* W/2 annotat* ) ) ) )
\end{quote}

\subsection*{Web of Science (04/15/24)}

\begin{quote}
((natural NEAR/2 language NEAR/2 process*) OR (large NEAR/2 language NEAR/2 model*) OR "NLP" OR "LLM" OR ((data OR text) NEAR/2 (mine* OR "mining")) OR "named entity" OR semantic) (Abstract) \\
AND (genet* OR genom* OR genot* OR "sequencing" OR ((single NEAR/2 nucleotide NEAR/2 polymorph*) OR "SNP*")) (Abstract) \\
AND (token* OR "transformer" OR "bert" OR (regulat* NEAR/2 annotat*)) (Abstract)
\end{quote}

\subsection*{ACM (04/15/24)}

\begin{quote}
[[Abstract: "natural language processing"] OR [Abstract: or] OR [Abstract: "large language model"] OR [Abstract: or "nlp" or "llm" or] OR [Abstract: "data mining"] OR [Abstract: "text mining"] OR [Abstract: "named entity"] OR [Abstract: or semantic]] \\
AND [[Abstract: genet* OR genom* OR genot* OR] OR [Abstract: "single nucleotide polymorph"] OR [Abstract: or "snp" or "snps"]] \\
AND [Abstract: token* OR "transformer" OR "bert"]
\end{quote}

\end{appendices}

\end{document}

%% file: table/table1.r0.tex
\definecolor{lightblue}{rgb}{0.937, 0.961, 0.996}

\rowcolors{2}{}{lightblue}

\begin{longtable}{>{\raggedright}p{1.5cm}>{\raggedright}p{1.5cm}>{\raggedright}p{3cm}>{\raggedright}p{1.5cm}>{\raggedright}p{4cm}>{\raggedright}p{1cm}>{\raggedright}p{1.5cm}>{\raggedright}p{2cm}>{\raggedright}p{2cm}>{\raggedright\arraybackslash}p{1.5cm}}
\caption{An overview of NLP application using big genomic data} \label{tab:nlp_genomic}\\

\toprule
Ref. & Model & Goal \& Aim & Data & Sample Size & Parameter Size & Tokenization & Transformer Architecture & Model Derived & Data\textbackslash Model Availability\\ 
\midrule
\endfirsthead

\multicolumn{9}{l}%
{{\bfseries \tablename\ \thetable{} -- continued from previous page}} \\
\toprule
Ref. & Model & Goal \& Aim & Data & Sample Size & Parameter Size & Tokenization & Transformer Architecture & Model Derived & Data\textbackslash Model Availability\\ 
\midrule
\endhead

\rowcolor{white}
\midrule 
\rowcolor{white}\multicolumn{10}{r}{Continued on next page} \\
\endfoot

\bottomrule
\endlastfoot
\citet{212} & Transformer-XL plus enhancement & Sequence labeling tasks & DNA sequences & 9283304 samples & 185346 to 462402 & Sequential segments of 512 nucleotides & Transformer-XL + convolutional layer & Pretrain & Yes\textbackslash Yes\\

\citet{1045} & DistilBERT + CRF + Attention Mask & Detect CpG islands in DNA sequences
; Promoter prediction; epigenetic causes identification & DNA sequences & 233,004 sequences\textsuperscript{a} & 66M & BPE & DistilBERT & Pretrain \& Fine tune (CpG island detection) & Yes\textbackslash No\\

\citet{1101} & DNABERT & Capture understanding of DNA sequences; Predict regulatory elements & DNA sequences & 690 TF ChIP-seq profiles & 110M & K-mer (3, 4, 5, 6) & BERT & Pretrain \& Fine tune & Yes\textbackslash Yes\\

\citet{232} & BERT-CNN & Identify DNA enhancers & DNA sequences & 1484 each (training); 200 each (test) & 1,317,442 & Nucleotides as words, DNA sequences as sentences & BERT + 2D CNN & Pretrain \& Fine tune & Yes\textbackslash Yes\\

\citet{190} & BERT-Promoter & Improve the prediction of DNA promoters and their strength & DNA sequences & 3382 each\textsuperscript{b} & 110M & DNA sequences split into 81-bp fragments & BERT + SHAP & Pretrain & Yes\textbackslash Yes\\

\citet{174} & TFBERT & Improve the prediction of DNA--protein binding sites & DNA sequences & 690 ChIP-seq datasets\textsuperscript{c} & 110M & K-mer & BERT & Pretrain \& Fine tune & Yes\textbackslash Yes\\

\citet{1064} & DeepViFi & Detect Oncoviral Infections in Cancer Genomes using Transformers & DNA sequences & 1,145,800 reads & 8 encoder & Each base-pair as a token & Self-attention heads & Pretrain & Yes\textbackslash Yes\\

\citet{902} & GENEMASK-based DNABERT & Improve MLM training efficiency for gene sequences & DNA sequences & Prom-core \& Prom-300: 53,276 training, 5,920 test; Splice-40: 24,300 training, 3,000 test; Cohn-enh: 20,843 training, 6,948 test & 110M & k-mer (k=6) & BERT & Genomic specific pretrain paradigm & Yes\textbackslash Yes\\

\citet{714} & BERT-5mC & Predict 5mC sites of DNA & DNA sequences & Training: 55,800 positive, 658,861 negative; Testing: 13,950 positive, 164,715 negative & NR\textsuperscript{d} & K-mer (k=3) & BERT & Pretrain \& Fine tune & Yes\textbackslash Yes\\

\citet{176} & SMFM & Identify and characterize DNA enhancers & DNA sequences & 2968 samples (1484 enhancers and 1484 non-enhancers) & NR & K-mer (k=3) & BERT & Fine tune & Yes\textbackslash Yes\\

\citet{1024} & DNABERT & Predict Protein-DNA binding sites & DNA sequences & 45 public transcription factor ChIP-seq datasets with DNA sequence samples of 101 bp & 110M & K-mer (k=6) & BERT; multi-headed self-attention & Fine tune & Yes\textbackslash No\\

\citet{979} & SemanticCAP & Chromatin accessibility prediction & DNA sequence & MT 244692, PC 418624; PH 503816, NO 264264; MU 266868, NP 283148 & 5.61M & Character based inputs & BERT & Pretrain & Yes\textbackslash Yes\\

\citet{1000} & moDNA & Genome embedding; Promoter prediction; Transcription factor binding sites prediction & Non-coding DNA sequences & NR & NR & 6-mers & BERT & Pretrain on human genome data \& Fine tune & Yes\textbackslash No\\

% Cui, 2024\cite{308} & scGPT & Sequence labeling tasks & scRNA-seq; scATAC-seq; multi-omic data & Over 33 million & NR & GeneNames BinnedValues & Multi-Head Attention & Pretrain \& Fine tune \\

\citet{946}& BERT + CRF & Detect CpG islands in DNA sequences; Promoter prediction; epigenetic causes identification & DNA sequences with annotated CpG islands & 233,004 sequences\textsuperscript{a} & 110M & BPE & BERT with CRF layer & Pretrain \& Fine tune (CpG island detection) & Yes\textbackslash No\\

% Khan, 2023\cite{882} & scBERT & Detect reusability \& generalizability; Cell-type annotation in single-cell RNA sequencing data & Single-cell RNA sequences & 23,911 cells & 110M & gene2vec & BERT & Evaluate on a different dataset \& distribution \\

\citet{507} & Transformer DeepLncLoc & Predict the abundance of mRNA (gene expression levels) & DNA sequences; mRNA half-life features; Transcription factors & 18,000 gene sequences with their expression values & 123,881 & K-mer (k=3) & Vanilla encoder block + DeepLncLoc embedding & Evaluate and use the output embedding & Yes\textbackslash Yes\\

\citet{656} & MuLan-Methyl & Predict DNA methylation sites N6-adenine, N4-cytosine, and 5-hydroxymethylcytosine & DNA sequences; Taxonomy lineages & 250,599 samples across 12 genomes & 110M\textsuperscript{e} & Custom Tokenizer\textsuperscript{f} & BERT; DistilBERT; ALBERT & Pretrain \& Fine tune & Yes\textbackslash Yes\\

\citet{48} & MSCAN & Identify RNA methylation sites & RNA sequences & m1A\_train0: 593 positive, 5930 negative; m6A\_train: 41,307 positive, 41,307 negative & NR & Word2vec embedding k-mer (k=3) & Multi-scale self- and cross-attention mechanisms with multi-head attention & Pretrain \& Fine tune & Yes\textbackslash No\\

\citet{859} & MTTLm6A & Predict base-resolution mRNA m6A sites & RNA sequences & m1A sites: 1987 positive samples, 2249 negative samples (Homo sapiens); m6A sites: 24669 m6A sites (S. cerevisiae) & NR & One-hot encoding & CNN; Multi-head attention & Fine tune & No\textbackslash No\\

\citet{69} & BCMCMI & Predict potential circRNA-miRNA interactions & circRNA and miRNA sequences & circBank: 9589 (2115 circRNAs, 821 miRNAs) CMI-9905: 9905 (2346 circRNAs, 926 miRNAs) & NR & BERT-based tokenization with WordPiece embeddings & BERT & Directly use BERT to get embedding & Yes\textbackslash Yes\\

\citet{12} & BioDGW-CMI & Predict circRNA-miRNA interactions & RNA sequences; Network structure & CMI-9905: 9905 (2346 circRNAs, 962 miRNAs); 
CMI-9589: 9589;
CMI-753: 753 & NR & K-mer (k=2 for miRNA and k=5 for circRNA) & BERT & Use existing pretrained model & Yes\textbackslash Yes\\

\citet{40} & miTDS & Predict miRNA-mRNA interactions & miRNA and mRNA sequences & 10 test datasets, each with 548 positive and 548 negative miRNA-mRNA pairs & 110M & BERT-based tokenization & BERT & Fine tune & Yes\textbackslash Yes\\

\citet{34} & TMSC-m7G & Predict RNA N7-methylguanosine (m7G) sites & RNA sequences with N7-methylguanosine modification sites & Benchmark: 741 positives, 741 negatives (balanced); Independent: 334 positives, 3340 negatives (imbalanced) & NR & K-mer, then
multi-sense-scaled word embedding & Transformer with CNN layer & Fine tune & Yes\textbackslash No\\

\citet{940} & Transformer-based DNA methylation detection model & Detect DNA methylation on ionic signals & Nanopore sequencing data & NR & NR & One-hot encoding & BERT & Fine tune & No\textbackslash Yes\\

\citet{986} & CGCD & Predict optimized potential anti-breast cancer therapeutic target genes & Multi-omics data & 105 breast cancer patients & 65M & Gene expression values as tokens & Transformer encoder & NR & Yes\textbackslash No\\

\citet{1027} & SETQUENCE, SETOMIC & Enhance tumor type classification; Provide ML model which can hand over omics data & Transcriptome expression data; Somatic mutation data & 544 healthy \& 7518 tumor samples across 32 cancer types & NR & 6-mers & DNABERT, DNN & Pretrain \& Fine tune & Yes\textbackslash Yes\\

\citet{143} & IGnet & Automated classification of Alzheimer\textquotesingle s disease & 3D MRI; SNP; CNV markers & ADNI-1 subset with 379 participants (174 AD patients and 205 normal controls) & NR & SNPs \{0,1, 2\}\textsuperscript{g} & 3D CNN for CV; two-layer transformer for genetic sequence & Train end to end & Yes\textbackslash No\\

\end{longtable}
\begin{compactitem}
\footnotesize
\item[a] 61,051 sequences containing 142,325 CpG islands 
\item[b] 3,382 promoters (1,591 strong and 1791 weak promoter samples) and 3,382 non-promoters 
\item[c]  4,153,122 training samples, 461,458 validation samples, 800,000 testing samples
\item[d] NR: Not Reported 
\item[e]  BERT: 110M; DistilBERT: 40\% of BERT; ALBERT: reduced size with cross-layer sharing 
\item[f]  A custom tokenizer that can capture any sample represented by 6-mer DNA words and a textual description of taxonomic lineage
\item[g]  SNPs encoded as {0,1,2}; selected with Fisher's test, concatenated with APOE 
\end{compactitem}